\documentclass{article}
\usepackage{amsmath}

\newcommand{\qed}{\nobreak \ifvmode \relax \else
      \ifdim\lastskip<1.5em \hskip-\lastskip
      \hskip1.5em plus0em minus0.5em \fi \nobreak
      \vrule height0.75em width0.5em depth0.25em\fi}
\newcommand{\beq}{\begin{equation}}
\newcommand{\eeq}{\end{equation}}

\def\opone{\leavevmode\hbox{\small1\kern-3.8pt\normalsize1}}

\input{tcilatex}
\begin{document}

\centerline{\large\bf Uncertainty relation of mixed states by means
of Wigner-Yanase-Dyson information }

\centerline{D. Li$^{a}$\footnote{email
address:dli@math.tsinghua.edu.cn}, X. Li$^{b}$, F. Wang$^{c}$, H.
Huang$^{d}$, X. Li$^{e}$, L. C. Kwek$^{f}$ }

\centerline{$^a$ Dept of mathematical sciences, Tsinghua University,
Beijing 100084 CHINA}

\centerline{$^b$ Department of Mathematics, University of
California, Irvine, CA 92697-3875, USA}

\centerline{$^c$ Insurance Department, Central University of Finance
and Economics, Beijing 100081, CHINA}

\centerline{$^d$ Electrical Engineering and Computer Science Department} %
\centerline{ University of Michigan, Ann Arbor, MI 48109, USA}

\centerline{$^e$ Dept. of Computer Science, Wayne State University,
Detroit, MI 48202, USA}

\centerline{$^f$ National Institute of
      Education,
      Nanyang Technological University, 1 Nanyang Walk, Singapore
      637616} 
\centerline{Centre for Quantum Technologies, National University of
Singapore, 3 Science Drive 2, Singapore 117543} 
\centerline{
Institute of Advanced Studies (IAS), Nanyang Technological
University, 60 Nanyang View Singapore 639673}

Abstract

The variance of an observable in a quantum state is usually used to describe
Heisenberg uncertainty relation. For mixed states, the variance includes
quantum uncertainty and classical uncertainty. By means of the skew
information and the decomposition of the variance, a stronger uncertainty
relation was presented by Luo in [Phys. Rev. A 72, 042110 (2005)]. In this
paper, by using Wigner-Yanase-Dyson information which is a generalization of
the skew information, we propose a general uncertainty relation of mixed
states.

PACS {03.65.Ta}

Keywords: Heisenberg uncertainty relation, the skew information, Dyson
information

\section{Introduction}

In quantum measurement theory, the Heisenberg uncertainty principle provides
a fundamental limit for the measurements of incompatible observables. On the
other hand, as dictated by Cramer-Rao's lower bound, there is also an
ultimate limit for the resolution of any unbiased parameter (see for
instance, \cite{jarda}), and this lower bound is given by a quantity called
Fisher information. A long time ago, Wigner demonstrated that it is more
difficult to measure observables that do not commute with some additive
conserved quantity. Thus, observables not commuting with some conserved
quantity cannot be measured exactly and only approximate measurement is
possible. This trade-off in measurement forms the basis of the well-known
Wigner-Araki-Yanase theorem. In their study of quantum measurement theory,
Wigner and Yanase introduced a quantity called the skew information. As
shown in \cite{Luo3}, the skew information is essentially a form of Fisher
information.

The skew information for a mixed state $\rho $ relative to a self-adjoint
``observable", $A$, is defined as $I(\rho $, $A)=$ $-\frac{1}{2} 
\mbox{\rm
Tr} \rho ^{1/2}$, $A]^{2}$. This definition was subsequently generalized by
Dyson as $I_{\alpha }(\rho ,A)=-\frac{1}{2}\mbox{\rm Tr} ([\rho ^{\alpha },
A][\rho ^{1-\alpha }, A])$, where $0<\alpha <1$ \cite{Wigner}. When $\alpha
=1/2$, $I_{\alpha }(\rho $, $X)$ is reduced to the skew information. The
convexity of $I_{\alpha }(\rho ,A)$ was finally resolved by Lieb\cite%
{Lieb,Lieb2}.

The von Neumann entropy of $\rho $, defined as $S(\rho )=-tr\rho \ln \rho $,
has been widely used as a measure of the uncertainty of a mixed state. This
quantity, profoundly rooted in quantum statistical mechanics, possesses
several remarkable and satisfactory properties. Like all measures, the von
Neumann entropy, together with its classical analog called the Shannon
entropy, is not always the best measure under certain contexts. In \cite%
{Luo1,Luo2,Luo3,Luo4}, the skew information was proposed as means to unify
the study of Heisenberg uncertainty relation for mixed states.

It is well know in the standard textbooks\ that the Heisenberg uncertainty
relation for any two self-adjoint operators $X$ and $Y$ is given by 
\begin{equation}
V(\rho ,X)V(\rho ,Y)\geq \frac{1}{4}||\mbox{\rm Tr}(\rho \lbrack X,Y]||^{2}.
\label{Heisenberg}
\end{equation}%
Note that $[$,$]$ is commutator, i.e. $[A$, $B]=AB-BA$ and the variance of
the observable $X$ with respect to $\rho $ is 
\begin{equation}
V(\rho ,X)=\mbox{\rm Tr}(\rho X^{2})-(\mbox{\rm Tr}(\rho X))^{2}.
\label{variance}
\end{equation}%
A similar definition applies to $V(\rho ,Y)$.

When $\rho $ is a mixed state, Luo showed that the variance comprises of two
terms: a quantum uncertainty term and a classical uncertainty term\cite%
{Luo1,Luo2}. He separated the variance into its quantum and classical part
by using the skew information. He interpreted $I(\rho $, $X)$ as the quantum
uncertainty of $X$ in $\rho $ by the Bohr complementary principle and $%
V(\rho ,X)-I(\rho $, $X)$ as the classical uncertainty of the mixed state.
He then considered $U(\rho ,X)=\sqrt{V^{2}(\rho ,X)-[V(\rho ,X)-I(\rho
,X)]^{2}}$ as a measure of quantum uncertainty. Thus, he obtained the
following two inequalities for the uncertainty relation. 
\begin{equation}
I(\rho ,X)J(\rho ,Y)\geq \frac{1}{4}||\mbox{\rm Tr} (\rho \lbrack X,Y]||^{2}.
\label{Luo}
\end{equation}
\begin{equation}
U(\rho ,X)U(\rho ,Y)\geq \frac{1}{4}||\mbox{\rm Tr} (\rho \lbrack X,Y]||^{2}.
\label{Luo2}
\end{equation}%
where $J(\rho $, $Y)=\frac{1}{2}\mbox{\rm Tr} \{\rho ^{1/2}$, $Y_{0}\}^{2}$%
,\ and $Y_{0}=Y-\mbox{\rm Tr} (\rho Y)$. The notation \ $\{$ $\}$ is the
anticommutator, i.e. $\{A$, $B\}=AB+BA$.

This article is organized as follows: In section \ref{sec2}, we discuss
various properties of the Wigner-Yanase-Dyson information. We show using a
counter example that it need not satisfy the uncertainty relation obtained
from the skew information. In section \ref{sec3}, we formulate an
uncertainty relation for Wigner-Yanase-Dyson information. Finally, in
section \ref{sec4}, we reiterate our main results. We have also provided two
appendices concerning the proof of the new uncertainty principle and
additivity of the Wigner-Yanase-Dyson information.

\section{Wigner-Yanase-Dyson information violates Heisenberg uncertainty
relation}

\label{sec2}

In this paper, we extend the above discussion to Wigner-Yanase-Dyson
information. The skew information proposed by Dyson can also be written as 
\begin{eqnarray}
I_{\alpha }(\rho ,X) &=&\mbox{\rm Tr} (\rho X^{2})-\mbox{\rm Tr} (\rho
^{\alpha }X\rho ^{1-\alpha }X)  \notag \\
&=&\mbox{\rm Tr} (\rho X_{0}^{2})-\mbox{\rm Tr} (\rho ^{\alpha }X_{0}\rho
^{1-\alpha }X_{0})\text{, }  \label{eq1}
\end{eqnarray}%
where $X_{0}=X-\mbox{\rm Tr} (\rho X)$. $I_{\alpha }(\rho ,X)$ is positive
from Eq. (\ref{q-info-2}). Similarly, we define $J_{\alpha }(\rho ,Y)=\frac{1%
}{2}tr(\{\rho ^{\alpha }$, $Y_{0}\}\{\rho ^{1-\alpha }$, $Y_{0}\})$. When $%
\alpha =1/2$, $J_{\alpha }(\rho $, $Y)$ is reduced to $J(\rho $, $Y)$. As
well, we can define $J_{\alpha }(\rho ,X)$, $J_{\alpha }(\rho ,A)$, and $%
J_{\alpha }(\rho ,B)$. By calculating,

\begin{eqnarray}
&&J_{\alpha }(\rho ,Y)=  \notag \\
&& \mbox{\rm Tr} (\rho Y_{0}^{2})+\mbox{\rm Tr} (\rho ^{\alpha }Y_{0}\rho
^{1-\alpha }Y_{0})=  \notag \\
&&\mbox{\rm Tr} (\rho Y^{2})+\mbox{\rm Tr} (\rho ^{\alpha }Y\rho ^{1-\alpha
}Y)-2(\mbox{\rm Tr} \rho Y)^{2}.  \label{eq2}
\end{eqnarray}%
$J_{\alpha }(\rho $, $Y)$ is also positive from Eq. (\ref{ineq4}) in this
paper.

Adopting the Luo's interpretations, by the following properties of
Wigner-Yanase-Dyson information we interpret $I_{\alpha }(\rho ,X) $ as
quantum uncertainty of $X$ in $\rho $, $V(\rho ,X)-I_{\alpha }(\rho $, $X)$
as the classical mixing uncertainty, and $U_{\alpha }(\rho ,X)=\sqrt{%
V^{2}(\rho ,X)-[V(\rho ,X)-I_{\alpha }(\rho ,X)]^{2}}$ as a measure of
quantum uncertainty. \ Lieb studied the properties of Wigner-Yanase-Dyson
information in \cite{Lieb}.\ Wigner-Yanase-Dyson information satisfies the
following requirements.

(1). Wigner-Yanase-Dyson conjecture about the convexity of $I_{\alpha }(\rho
,X)$ with respect to $\rho $ was proved by Lieb \cite{Lieb}.

(2). Wigner-Yanase-Dyson information $I_{\alpha }(\rho ,X)$ is additive
under the following sense (See \cite{Luo3} and \cite{Lieb}). Let $\rho _{1}$
and $\rho _{2}$ be two density operators of two subsystems, and $A_{1}$
(resp. $A_{2}$) be a self-adjoint operator on $H^{1}$ (resp. $H^{2}$). $%
I_{\alpha }(\rho ,X)$ is additive if $I_{\alpha }(\rho _{1}\otimes \rho _{2}$%
, $A_{1}\otimes I_{2}+I_{1}\otimes A_{2})=I_{\alpha }(\rho _{1}$, $%
A_{1})+I_{\alpha }(\rho _{2}$, $A_{2})$, where $I_{1}$ and $I_{2}$ are the
identity operators for the first and second systems, respectively. For the
proof see Appendix B.

(3). $J_{\alpha }(\rho $, $Y)$ is also additive under the above sense. For
the proof see Appendix B.

(4). However, Hansen showed that Wigner-Yanase-Dyson information is not
subadditive \cite{Hansen}. For the definition of subadditivity see \cite%
{Lieb} and \cite{Hansen}.

(5). $J_{\alpha }(\rho $, $Y)$ is concave with respect to $\rho $. This is
because $tr(\rho Y_{0}^{2})$ is linear operator with respect to $\rho $ and $%
tr(\rho ^{\alpha }Y_{0}\rho ^{1-\alpha }Y_{0})$ is concave with respect to $%
\rho $.

(6). When $\rho $ is pure, $V(\rho ,X)=I_{\alpha }(\rho $, $X)$. Thus,
Wigner-Yahase-Dyson information reduces to the variance. That is, the
variance $V(\rho ,X)$ does not include the classical mixing uncertainty
because of no mixing. In other words, the variance only includes the quantum
uncertainty of $X$ in $\rho $. The case in which $\alpha =1/2$ was discussed
in \cite{Luo2}. \ 

The above fact can be argued as follows. When $\rho $ is pure, $tr(\rho
^{\alpha }X_{0}\rho ^{1-\alpha }X_{0})=$ $(tr(\rho X_{0}))^{2}=0$.$\ $Thus, $%
I_{\alpha }(\rho ,X)=tr(\rho X_{0}^{2})=V(\rho ,X)$.

(7). When $\rho $ is a mixed state, $V(\rho ,X)\geq I_{\alpha }(\rho $, $X)$%
. This is because $tr(\rho ^{\alpha }X\rho ^{1-\alpha }X)$ $=$ $tr((\rho
^{\alpha /2}X\rho ^{(1-\alpha )/2})$ $(\rho ^{\alpha /2}X\rho ^{(1-\alpha
)/2})^{\dagger })\geq 0$. Also, see Eq. (\ref{I-2}) in this paper. The case
in which $\alpha =1/2$ was discussed in \cite{Luo2}. \ 

(8). When $\rho $ and $A$ commute, according to the discussion for the skew
information in \cite{Luo1,Luo4}, the quantum uncertainty should vanish and
thus, the variance only includes the classical uncertainty. We can argue
that the above conclusion is also true for Wigner-Yanase-Dyson information.
When $\rho $ and $A$ commute, it is well known that $\rho $ and $A$ have the
same orthonormal eigenvector basis \cite{Mika}. Hence, $\rho ^{\alpha }$ and 
$A$ also commute. By the definition in Eq. (\ref{eq1}), Wigner-Yanase-Dyson
information $I_{\alpha }(\rho ,X)$ vanishes.

However, $I_{\alpha }(\rho ,X)$\ and $J_{\alpha }(\rho $, $Y)$\ do not
satisfy Eq. (\ref{Luo}).\ We give the following counter example\textbf{\ }%
for Eq. (\ref{Luo}).

Let $n=2$, $\alpha =1/4$, and $\rho $ have the eigenvalues $\lambda _{1}=1/4$
and $\lambda _{2}=3/4$. Since $A$ and $B$ are self-adjoint, then we write $%
A=\left( 
\begin{tabular}{ll}
$x$ & $u+iv$ \\ 
$u-iv$ & $y$%
\end{tabular}%
\right) $, $B=\left( 
\begin{tabular}{ll}
$a$ & $c+di$ \\ 
$c-di$ & $b$%
\end{tabular}%
\right) $. In this example, $u=4$, $v=2$, $a=b=0$, $c=1$, and $d=-5$. By
calculating $I_{\alpha }(\rho ,A)$ in Eq. (\ref{q-info-2}) and $J_{\alpha
}(\rho ,B)$ in Eq. (\ref{eq5}), $I_{\alpha }(\rho ,A)J_{\alpha }(\rho
,B)=[1-(\lambda _{1}^{\alpha }\lambda _{2}^{1-\alpha }+\lambda _{2}^{\alpha
}\lambda _{1}^{1-\alpha })^{2}](u^{2}+v^{2})(c^{2}+d^{2})=99.83$. By
calculating $\mbox{\rm Tr}(\rho \lbrack A,B]$ in Eq. (\ref{eq7}), $\frac{1}{4%
}|\mbox{\rm Tr}(\rho \lbrack A,B]|^{2}=(\lambda _{1}-\lambda
_{2})^{2}(cv-du)^{2}=121$. Hence, it violates Eq. (\ref{Luo}). It implies
that the bound on the right side of the inequality in Eq. (\ref{Luo}) is too
large in this example. We need to get the appropriate lower bound for
Wigner-Yanase-Dyson information, i.e., we need to modify the term on RHS of
the inequality.

\section{The general uncertainty relation}

\label{sec3}

We replace $\mbox{\rm Tr} (\rho \lbrack X,Y]$ with $l_{\alpha }(\rho $, $X$, 
$Y)$ which is defined as follows: 
\begin{equation}
l_{\alpha }(\rho ,X,Y)=\mbox{\rm Tr} (\rho \lbrack X,Y])-\mbox{\rm Tr} \rho
^{\left\vert 2\alpha -1\right\vert }[X,Y].  \label{def3}
\end{equation}%
When $\alpha =1/2$, $l_{\alpha }(\rho $, $X$, $Y)$ reduces to $\mbox{\rm Tr}
(\rho \lbrack X$, $Y])$. In \cite{Luo1}, Luo\ defined $k=\mathrm{i}[\rho
^{1/2}$, $X_{0}]t+\{\rho ^{1/2},Y_{0}\}$, where $t\in R$ and $\mathrm{i}$ is
an imaginary number. From $\mbox{\rm Tr} (kk^{\dagger })\geq 0$, by
expanding $\mbox{\rm Tr} (kk^{\dagger })$, he derived $\mbox{\rm Tr}
(kk^{\dagger })=2(I[\rho $, $X]t^{2}+\mathrm{i}(tr(\rho \lbrack X$, $%
Y])t+J[\rho $, $Y])\geq 0$. Since the above inequality is true for any real $%
t$, Luo obtained the inequality in Eq. (\ref{Luo}). However, unlike his
previous case, the form of $I_{\alpha }(\rho ,X)$ does not allow us to
employ the trick $k=\mathrm{i}[\rho ^{\alpha }$, $X_{0}]t+\{\rho ^{\alpha
},Y_{0}\}$ nor $k=\mathrm{i}[\rho ^{1-\alpha }$, $X_{0}]t+\{\rho^{1-\alpha
},Y_{0}\}$ to derive the uncertainty relation from $\mbox{\rm Tr}
(kk^{\dagger })\geq 0$. The proof becomes more involved and one needs to
modify the RHS of the previous uncertainty relation.

In Appendix A, we see that if $A$ and $B$ are self-adjoint observables, then 
\begin{equation}
I_{\alpha }(\rho ,A)J_{\alpha }(\rho ,B)\geq \frac{1}{4}||l_{\alpha }(\rho
,A,B)||^{2}\text{, }  \label{uncertainty-3}
\end{equation}
and 
\begin{equation}
I_{\alpha }(\rho ,B)J_{\alpha }(\rho ,A)\geq \frac{1}{4}||l_{\alpha }(\rho
,A,B)||^{2}\text{. }  \label{uncertainty-4}
\end{equation}
If we denote $U_{\alpha }(\rho, \mathcal{O})$ as $\sqrt{V^{2}(\rho, \mathcal{%
O} )-[V(\rho, \mathcal{O})-I_{\alpha }(\rho ,\mathcal{O})]^{2}}$, we see
that by Eq. (\ref{variance}) and Eq.(\ref{eq1}) (and the analogous form for $%
J_\alpha(\rho,\mathcal{O})$), $U_{\alpha }(\rho , \mathcal{O})=\sqrt{%
I_{\alpha }(\rho ,\mathcal{O})J_{\alpha }(\rho ,\mathcal{O})}$, where $%
\mathcal{O}$ is either the operator $A$ or $B$. Thus, we obtain our main
result from Eqs. (\ref{uncertainty-3}) and (\ref{uncertainty-4}), 
\begin{equation}
U_{\alpha }(\rho ,A)U_{\alpha }(\rho ,B)\geq \frac{1}{4}||l_{\alpha }(\rho
,X,Y)||^{2}.
\end{equation}

For the counter example in Sec. \ref{sec2}, a direct calculation of Eq. (\ref%
{eq6}) yields $\frac{1}{4}||l_{\alpha }(\rho ,A,B)||^{2}=$ $8.\,\allowbreak
687\,4$. Therefore, the inequality in Eq. (\ref{uncertainty-3}) holds in
this case.

\section{Summary}

\label{sec4}

In \cite{Luo1}, Luo presented a refined Heisenberg uncertainty relation. In
this paper, we demonstrate some properties of Wigner-Yanase-Dyson
information and provide a counter example to show that Wigner-Yanase-Dyson
information does not in general satisfy Heisenberg uncertainty relation. We
have also proposed a new general uncertainty relation of mixed states based
on Wigner-Yanase-Dyson information. Bell-type inequalities based on the skew
information have been proposed as nonlinear entanglement witnesses \cite%
{chen}. We note here that similar Bell-type inequalities with the advantage
of an additional $\alpha$ parameter for fine adjustments could also be
constructed from the uncertainty principle derived from the
Wigner-Yanase-Dyson information.

\section*{Appendix A. Proof of uncertainty relation}

\setcounter{equation}{0} \renewcommand{\theequation}{A\arabic{equation}}

By spectral decomposition,\ there exists an orthonormal basis $\{x_{1}$,..., 
$x_{n}\}$\ consisting of eigenvectors of $\rho $. Let $\lambda _{1}$, ..., $%
\lambda _{n}$ be the corresponding eigenvalues, where $\lambda
_{1}+...+\lambda _{n}=1$ and $\lambda _{i}\geq 0$. Thus, $\rho $ has a
spectral representation 
\begin{equation}
\rho =\lambda _{1}|x_{1}\rangle \langle x_{1}|+....+\lambda
_{n}|x_{n}\rangle \langle x_{n}|.  \label{spectral-1}
\end{equation}


\subsection*{\textit{1. Calculating }$I_{\protect\alpha }(\protect\rho $%
\textit{, }$A)$}

By Eq. (\ref{spectral-1}), $\rho A^{2}=\lambda _{1}|x_{1}\rangle \langle
x_{1}|A^{2}+....+\lambda _{n}|x_{n}\rangle \langle x_{n}|A^{2}$ and 
\begin{eqnarray}
\mbox{\rm Tr}\rho A^{2} &=&\lambda _{1}\langle x_{1}|A^{2}|x_{1}\rangle
+....+\lambda _{n}\langle x_{n}|A^{2}|x_{n}\rangle   \notag \\
&=&\lambda _{1}||A|x_{1}||^{2}+....+\lambda _{n}||A|x_{n}||^{2}.  \label{I-1}
\end{eqnarray}

Moreover, since $\rho ^{\alpha }A=\lambda _{1}^{\alpha }|x_{1}\rangle
\langle x_{1}|A+....+\lambda _{n}^{\alpha }|x_{n}\rangle \langle x_{n}|A$
and $\rho ^{1-\alpha }A=\lambda _{1}^{1-\alpha }|x_{1}\rangle \langle
x_{1}|A+....+\lambda _{n}^{1-\alpha }|x_{n}\rangle \langle x_{n}|A$, we
have, $\rho ^{\alpha }A\rho ^{1-\alpha }A=\sum_{i,j=1}\lambda _{i}^{\alpha
}\lambda _{j}^{1-\alpha }|x_{i}\rangle \langle x_{i}|A|x_{j}\rangle \langle
x_{j}|A$. Thus 
\begin{eqnarray}
\mbox{\rm Tr} \rho ^{\alpha }A\rho ^{1-\alpha }A & = & \sum_{i,j=1}\lambda
_{i}^{\alpha }\lambda _{j}^{1-\alpha }\langle x_{i}|A|x_{j}\rangle \langle
x_{j}|A|x_{i}\rangle  \notag \\
& = & \sum_{i,j=1}\lambda _{i}^{\alpha }\lambda _{j}^{1-\alpha }||\langle
x_{i}|A|x_{j}\rangle ||^{2}.  \label{I-2}
\end{eqnarray}

From Eqs. (\ref{eq1}), (\ref{I-1}) and (\ref{I-2}), 
\begin{equation}
I_{\alpha }(\rho ,A)=\sum_{i=1}\lambda
_{i}||A|x_{i}||^{2}-\sum_{i,j=1}\lambda _{i}^{\alpha }\lambda _{j}^{1-\alpha
}||\langle x_{i}|A|x_{j}\rangle ||^{2}.  \label{q-info-1}
\end{equation}

Let $A=\{A_{ij}\}$ (resp. $B=\{B_{ij}\}$) be the matrix representation of
the operator $A$ (resp. $B$) corresponding to the orthonormal basis $\{x_{1}$%
,..., $x_{n}\}$. Then $\langle x_{i}|A|x_{j}\rangle =A_{ij}$, and 
\begin{eqnarray}
I_{\alpha }(\rho ,A) & = & \sum_{i\neq j}(\lambda _{i}-\lambda _{i}^{\alpha
}\lambda _{j}^{1-\alpha })\left\vert \left\vert A_{ij}\right\vert
\right\vert ^{2}  \notag \\
& = &\sum_{i<j}(\lambda _{i}+\lambda _{j}-\lambda _{i}^{\alpha }\lambda
_{j}^{1-\alpha }-\lambda _{i}^{1-\alpha }\lambda _{j}^{\alpha })\left\vert
\left\vert A_{ij}\right\vert \right\vert ^{2}\text{.}  \notag \\
& &  \label{q-info-2}
\end{eqnarray}


\subsection*{2. \textit{Calculating }$J_{\protect\alpha }(\protect\rho ,B)$}

Similarly, from Eqs. (\ref{eq2}) and (\ref{spectral-1}), we can obtain 
\begin{eqnarray}
J_{\alpha }(\rho ,B) & = & \sum_{i=1}\lambda
_{i}||B|x_{i}||^{2}+\sum_{i,j=1}\lambda _{i}^{\alpha }\lambda _{j}^{1-\alpha
}||\langle x_{i}|B|x_{j}\rangle ||^{2}  \notag \\
& & -2(\sum \lambda _{i}\langle x_{i}|B|x_{i}\rangle )^{2}.  \label{J-1}
\end{eqnarray}

Let $\langle x_{i}|B|x_{j}\rangle =B_{ij}$. Then, from Eq. (\ref{J-1}), 
\begin{eqnarray}
J_{\alpha }(\rho ,B) &=&2\sum_{i=1}\lambda _{i}\left\vert B_{ii}\right\vert
^{2}-2(\sum_{i=1}\lambda _{i}B_{ii})^{2}  \notag \\
&&+\sum_{i\neq j}(\lambda _{i}+\lambda _{i}^{\alpha }\lambda _{j}^{1-\alpha
})\left\vert \left\vert B_{ij}\right\vert \right\vert ^{2}\text{.}
\label{eq4}
\end{eqnarray}%
By simplifying, \ 
\begin{eqnarray}
J_{\alpha }(\rho ,B) &=&2\sum_{i=1}\lambda _{i}\left\vert B_{ii}\right\vert
^{2}-2(\sum_{i=1}\lambda _{i}B_{ii})^{2}  \notag \\
&&+\sum_{i<j}(\lambda _{i}+\lambda _{j}+\lambda _{i}^{\alpha }\lambda
_{j}^{1-\alpha }+\lambda _{i}^{1-\alpha }\lambda _{j}^{\alpha })\left\vert
\left\vert B_{ij}\right\vert \right\vert ^{2}\text{.}  \label{eq5}
\end{eqnarray}%
Since $x^{2}$ is convex, $(\sum_{i=1}\lambda _{i}B_{ii})^{2}\leq
\sum_{i=1}\lambda _{i}\left\vert B_{ii}\right\vert ^{2}$. So from Eq. (\ref%
{eq5}), 
\begin{equation}
J_{\alpha }(\rho ,B)\geq \sum_{i<j}(\lambda _{i}+\lambda _{j}+\lambda
_{i}^{\alpha }\lambda _{j}^{1-\alpha }+\lambda _{i}^{1-\alpha }\lambda
_{j}^{\alpha })\left\vert \left\vert B_{ij}\right\vert \right\vert ^{2}\text{%
.}  \label{ineq4}
\end{equation}


\subsection*{3. \textit{Calculating} $l_{\protect\alpha }(\protect\rho $%
\textit{, }$A$\textit{, }$B)$}

First we calculate $\mbox{\rm Tr}(\rho \lbrack A$, $B])$. By Eq. (\ref%
{spectral-1}), $\rho \lbrack A,B]=\lambda _{1}|x_{1}\rangle \langle x_{1}|[A$%
, $B]+....+\lambda _{n}|x_{n}\rangle \langle x_{n}|[A$, $B]$ and $%
\mbox{\rm
Tr}(\rho \lbrack A$, $B])=\lambda _{1}\langle x_{1}|[A$, $B]|x_{1}\rangle
+....+\lambda _{n}\langle x_{n}|[A$, $B]|x_{n}\rangle $. It is well known
that $Re\langle x_{i}|[A$, $B]|x_{i}\rangle =0$ and $\langle x_{i}|[A$, $%
B]|x_{i}\rangle =\mathrm{i}(2Im\langle x_{i}|AB|x_{i}\rangle )$, where $%
\mathrm{i}$ is an imaginary number. Consequently, $\mbox{\rm Tr}(\rho
\lbrack A$, $B])=2\mathrm{i}(\lambda _{1}Im\langle x_{1}|AB|x_{1}\rangle
+....+\lambda _{n}Im\langle x_{n}|AB|x_{n}\rangle )$. Therefore we obtain 
\begin{eqnarray}
\mbox{\rm Tr}(\rho \lbrack A,B]) &=&2\mathrm{i}Im(\lambda _{1}\langle
x_{1}|AB|x_{1}\rangle +...+\lambda _{n}\langle x_{n}|AB|x_{n}\rangle ) 
\notag \\
&=&2\mathrm{i}Im\sum_{j\neq i}\lambda _{i}A_{ij}B_{ji}.  \label{eq3}
\end{eqnarray}

Note that in Eq. (\ref{eq3})$\ A_{ii}$ and $B_{ii}$ are real because $A$ and 
$B$ are self-adjoint.\ Since $A_{ij}B_{ji}=(A_{ji}B_{ij})^{\ast }$, $\Im
\sum_{j\neq i}\lambda _{i}A_{ij}B_{ji}=Im\sum_{i<j}(\lambda _{i}-\lambda
_{j})A_{ij}B_{ji}$. Thus, by simplifying, 
\begin{equation}
\mbox{\rm Tr}(\rho \lbrack A,B])=2\mathrm{i}Im\sum_{i<j}(\lambda
_{i}-\lambda _{j})A_{ij}B_{ji}.  \label{eq7}
\end{equation}%
Moreover, 
\begin{equation}
\mbox{\rm Tr}\rho ^{\left\vert 2\alpha -1\right\vert }[A,B]=2\mathrm{i}%
Im\sum_{i<j}(\lambda _{i}^{\left\vert 2\alpha -1\right\vert }-\lambda
_{j}^{\left\vert 2\alpha -1\right\vert })A_{ij}B_{ji}.  \label{eq8}
\end{equation}%
Hence, from Eqs. (\ref{def3}), (\ref{eq7}) and (\ref{eq8}), 
\begin{equation}
l_{\alpha }(\rho ,A,B)=2\mathrm{i}\sum_{i<j}(\lambda _{i}-\lambda
_{j}-(\lambda _{i}^{\left\vert 2\alpha -1\right\vert }-\lambda
_{j}^{\left\vert 2\alpha -1\right\vert }))Im(A_{ij}B_{ji}).  \label{eq6}
\end{equation}


\subsection*{4. \textit{The proof of the uncertainty relation}\protect%
\bigskip}

From Eqs. (\ref{q-info-2}), (\ref{ineq4}) and (\ref{eq6}), for Eq. (\ref%
{uncertainty-3})\ we need to show 
\begin{eqnarray}
& & [\sum_{i<j}(\lambda _{i}+\lambda _{j}-\lambda _{i}^{\alpha }\lambda
_{j}^{1-\alpha }-\lambda _{i}^{1-\alpha }\lambda _{j}^{\alpha })\left\vert
\left\vert A_{ij}\right\vert \right\vert ^{2}] [\sum_{i<j}(\lambda
_{i}+\lambda _{j}+\lambda _{i}^{\alpha }\lambda _{j}^{1-\alpha }+\lambda
_{i}^{1-\alpha }\lambda _{j}^{\alpha })\left\vert \left\vert
B_{ij}\right\vert \right\vert ^{2}]  \notag \\
&\geq & \{\sum_{i<j}[\lambda _{i}-\lambda _{j}-(\lambda _{i}^{\left\vert
2\alpha -1\right\vert }-\lambda _{j}^{\left\vert 2\alpha -1\right\vert
})]Im(A_{ij}B_{ji})\}^{2}.  \label{ineq-1}
\end{eqnarray}
\newline
It is easy to know $[Im(A_{ij}B_{ji})]^{2}\leq \left\vert \left\vert
A_{ij}\right\vert \right\vert ^{2}\left\vert \left\vert B_{ij}\right\vert
\right\vert ^{2}$. Note that $\lambda _{i}+\lambda _{j}-\lambda _{i}^{\alpha
}\lambda _{j}^{1-\alpha }-\lambda _{i}^{1-\alpha }\lambda _{j}^{\alpha
}=(\lambda _{i}^{\alpha }-\lambda _{j}^{\alpha })(\lambda _{i}^{1-\alpha
}-\lambda _{j}^{1-\alpha })\geq 0$. \ By the Cauchy-Schwartz inequality, the
LHS of the inequality in Eq. (\ref{ineq-1}) $\geq \{\sum [(\lambda
_{i}+\lambda _{j})^{2}-(\lambda _{i}^{\alpha }\lambda _{j}^{1-\alpha
}+\lambda _{i}^{1-\alpha }\lambda _{j}^{\alpha
})^{2}]^{1/2}Im(A_{ij}B_{ji})\}^{2}$. Finally, what needs to be shown is 
\begin{eqnarray}
& & (\lambda _{i}+\lambda _{j})^{2}-(\lambda _{i}^{\alpha }\lambda
_{j}^{1-\alpha }+\lambda _{i}^{1-\alpha }\lambda _{j}^{\alpha })^{2}  \notag
\\
&\geq & |(\lambda _{i}-\lambda _{j})-(\lambda _{i}^{2\alpha -1}-\lambda
_{j}^{2\alpha -1})|^{2}\text{.}  \label{ineq-2}
\end{eqnarray}

\noindent It is easy to see that 
\begin{eqnarray*}
&&(\lambda _{i}+\lambda _{j})^{2}-(\lambda _{i}^{\alpha }\lambda
_{j}^{1-\alpha }+\lambda _{i}^{1-\alpha }\lambda _{j}^{\alpha })^{2} \\
&=&(\lambda _{i}-\lambda _{j})^{2}-(\lambda _{i}^{\alpha }\lambda
_{j}^{1-\alpha }-\lambda _{i}^{1-\alpha }\lambda _{j}^{\alpha })^{2}.
\end{eqnarray*}%
When $\alpha \geq 1/2$, 
\begin{eqnarray*}
&&(\lambda _{i}-\lambda _{j})^{2}-(\lambda _{i}^{\alpha }\lambda
_{j}^{1-\alpha }-\lambda _{i}^{1-\alpha }\lambda _{j}^{\alpha })^{2} \\
&=&(\lambda _{i}-\lambda _{j})^{2}-\lambda _{i}^{2(1-\alpha )}\lambda
_{j}^{2(1-\alpha )}(\lambda _{i}^{2\alpha -1}-\lambda _{j}^{2\alpha -1})^{2}
\\
&\geq &(\lambda _{i}-\lambda _{j})^{2}-(\lambda _{i}^{2\alpha -1}-\lambda
_{j}^{2\alpha -1})^{2} \\
&=&|(\lambda _{i}-\lambda _{j})-(\lambda _{i}^{2\alpha -1}-\lambda
_{j}^{2\alpha -1})| \\
&&\times \left\vert (\lambda _{i}-\lambda _{j})+(\lambda _{i}^{2\alpha
-1}-\lambda _{j}^{2\alpha -1})\right\vert  \\
&\geq &|(\lambda _{i}-\lambda _{j})-(\lambda _{i}^{2\alpha -1}-\lambda
_{j}^{2\alpha -1})|^{2}.
\end{eqnarray*}%
Note that the last inequality holds because $(\lambda _{i}-\lambda _{j})$
and $(\lambda _{i}^{2\alpha -1}-\lambda _{j}^{2\alpha -1})$ have the same
sign. Also, when $0<\alpha \leq 1/2$,\ we can prove the inequality in Eq. (%
\ref{ineq-2}) as follows: Let $\beta =1-\alpha $ with $1/2\leq \beta <1$.
Replacing $\alpha $ in Eq. (\ref{ineq-2}) with $1-$ $\beta $, we obtain $%
(\lambda _{i}+\lambda _{j})^{2}-(\lambda _{i}^{1-\beta }\lambda _{j}^{\beta
}+\lambda _{i}^{\beta }\lambda _{j}^{1-\beta })^{2}\geq |(\lambda
_{i}-\lambda _{j})-(\lambda _{i}^{2\beta -1}-\lambda _{j}^{2\beta -1})|^{2}$%
. This ends the proof.

\section*{Appendix B. Additivity}

\setcounter{equation}{0} \renewcommand{\theequation}{B\arabic{equation}}

The quantity $J_{\alpha }(\rho $, $B)$ is additive in the following sense: $%
J_{\alpha }(\rho _{1}\otimes \rho _{2}$, $B_{1}\otimes I_{2}+I_{1}\otimes
B_{2})=J_{\alpha }(\rho _{1}$, $B_{1})+J_{\alpha }(\rho _{2}$, $B_{2})$.
Using the notation in \cite{Lieb}, the proof proceeds by letting $\rho
_{12}=\rho _{1}\otimes \rho _{2}$ and $L=B_{1}\otimes I_{2}+I_{1}\otimes
B_{2}$. Setting $\rho _{12}^{\alpha }=\rho _{1}^{\alpha }\otimes \rho
_{2}^{\alpha }$, we have

$\rho _{12}^{\alpha }L\rho _{12}^{1-\alpha }L$ $=\rho _{1}^{\alpha
}B_{1}\rho _{1}^{1-\alpha }B_{1}\otimes \rho _{2}+\rho _{1}^{\alpha
}B_{1}\rho _{1}^{1-\alpha }\otimes \rho _{2}B_{2}$ $+\rho _{1}B_{1}\otimes
\rho _{2}^{\alpha }B_{2}\rho _{2}^{1-\alpha }+\rho \otimes \rho _{2}^{\alpha
}B_{2}\rho _{2}^{1-\alpha }B_{2}$, and 
\begin{eqnarray}
&&\mbox{\rm Tr}(\rho _{12}^{\alpha }L\rho _{12}^{1-\alpha }L)  \notag \\
&=&\mbox{\rm Tr}(\rho _{1}^{\alpha }B_{1}\rho _{1}^{1-\alpha }B_{1})+2%
\mbox{\rm Tr}(\rho _{1}B_{1})\mbox{\rm Tr}(\rho _{2}B_{2})+\mbox{\rm Tr}%
(\rho _{2}^{\alpha }B_{2}\rho _{2}^{1-\alpha }B_{2}).  \label{add1}
\end{eqnarray}%
Similarly, 
\begin{equation}
\mbox{\rm Tr}(\rho _{12}^{\alpha }L^{2})=\mbox{\rm Tr}(\rho _{1}B_{1}^{2})+2%
\mbox{\rm Tr}(\rho _{1}B_{1})\mbox{\rm Tr}(\rho _{2}B_{2})+\mbox{\rm Tr}%
(\rho _{2}B_{2}^{2}).  \label{add2}
\end{equation}%
From the above Eqs. (\ref{add1}) and (\ref{add2}), we can derive $I_{\alpha
}(\rho $, $B)$ is additive.

Similarly, 
\begin{equation}
\mbox{\rm Tr}(\rho _{12}^{\alpha }L)=\mbox{\rm Tr}(\rho _{1}B_{1})+\mbox{\rm
Tr}(\rho _{2}B_{2}).  \label{add3}
\end{equation}%
By Eqs. (\ref{add1}), (\ref{add2}), and (\ref{add3}), and the definition of $%
J_{\alpha }(\rho $, $B)$ in Eq. (\ref{eq2}), we can conclude that $J_{\alpha
}(\rho $, $B)$ is additive.

\textbf{Acknowledgments}

The first author wants to thank Prof. Jinwen Chen for his helpful discussion
about the inequality in Eq. (\ref{ineq-2}) and Mr Qin Zhang for the
discussion about the idea for $\mbox{\rm Tr} (KK^{\prime })\geq 0$. The
paper was supported by NSFC(Grants No.10875061, 60673034. KLC would like to
acknowledge financial support by the National Research Foundation \&
Ministry of Education, Singapore, for his visit and collaboration at
Tsinghua University.

\end{document}